\documentclass[journal]{IEEEtran}
\usepackage{subfigure}
\usepackage{cite}
\usepackage{citesort}
\usepackage{amsmath,epsfig,bm}
\usepackage{pifont}
\usepackage{paralist}
\usepackage{amssymb}
\usepackage{amsmath}
\usepackage{cancel}
\usepackage[normalem]{ulem}
\usepackage[dvips]{color}
\usepackage{algorithm}
\usepackage{etoolbox}
\patchcmd{\thebibliography}{\section*{\refname}}{}{}{}

\usepackage{pifont}
\usepackage{graphicx}
\usepackage{amssymb}
\usepackage{amsmath}
\usepackage{cancel}
\usepackage[normalem]{ulem}
\usepackage{algorithm}
\usepackage{algorithmic}

\begin{document}

\title{A Note on the Information-Theoretic-(in)Security of Fading Generated Secret Keys}

\author{Robert~Malaney$^*$

\thanks{$^*$Robert Malaney is with the School of
Electrical Engineering  and Telecommunications at the University
of New South Wales, Sydney, NSW 2052, Australia. email:
r.malaney@unsw.edu.au} }


\vspace{-3cm}

\maketitle
\begin{abstract}

In this work we explore the security of secret keys generated via the electromagnetic reciprocity of the wireless fading channel. Identifying a new sophisticated colluding attack, we explore the information-theoretic-security for such keys in the presence of an all-powerful adversary constrained only by the laws of quantum mechanics. Specifically, we calculate the reduction in the conditional mutual information between transmitter and receiver that can occur when an adversary with unlimited computational  and communication resources places directional-antenna interceptors at chosen locations. Such locations, in principal, can be arbitrarily  far from the intended receiver yet still influence the secret key rate.

\end{abstract}




The fast generation in real-world communication systems of an information-theoretic-secure
 key remains an ongoing endeavor. Currently, almost all   key generation systems being considered for commercial deployment scenarios are those based on the quantum mechanical properties of the information carriers.  However, although great progress has been made over the years with respect to quantum key distribution (QKD) systems, significant practical challenges remain before their deployment  becomes ubiquitous (see \cite{rev} for a recent review). The technical reasons for such a circumstance are many-fold, but they do give rise to a search for other sources of shared randomness that can be exploited for secret key generation. Transceivers connected via a wireless channel offer up one possibility  - via purely classical means.\footnote{Such a classical route is  particularly important since QKD directly over wireless channels in the GHz range is constrained to short distances \cite{weed1,neda}.}

 Indeed, for many years it has been known that the random fading inherent in the general wireless environment (coupled with electromagnetic reciprocity) is a potential source of secret key generation (for recent reviews see \cite{surv1,surv2,rev2,poor}). Conditioned on  the reasonable (and trivial to verify) assumption that Eve (the adversary)  is not in the immediate vicinity (a few cm at GHz frequencies) of Bob (the legitimate receiver), it is often stated that fading can lead to information-theoretic-secure keys. Here we clarify this is not the case when ${M_e} \to \infty$, $M_e$ being the number of receiving devices held by Eve.


An all-powerful adversary constrained only by nature herself is used in almost all security analyses of QKD systems \cite{rev} - and it is the protection from such an unearthly adversary that lends quantum-based key systems their acclaimed security status. Such acclaimed security is information-theoretic-secure (conditioned on the classical channel being authenticated) and remains in place irrespective of the computational resources or energy afforded to the adversary, even as ${M_e} \to \infty$.

Here, we explore an attack by such an all-powerful Eve on wireless fading generated key systems. More specifically, we quantify how this Eve can place   directional-antenna receivers (e.g. apertures,  linear-arrays, phased-arrays, etc.) at multiple locations in real-word scattering environments, and in principal drive to arbitrary low levels the conditional mutual information between Bob  and Alice (the transmitter). Practical scenarios invoking limited forms of the attack are discussed, showing (at the very least) how current fading-generated key systems are partially susceptible to the attack. 

We will take the term `information-theoretic-secure' to specifically mean the following. Conditioned on some well-defined assumption (or restriction) on the system model, but independent of  the capabilities of an adversary (other than being constrained by natural law), the key information accessible to an adversary can be driven to arbitrary small levels for increasing use of some system resource. Specifically, consider some series of observations of the random variables ${X} = \left( {{X_1},{X_2}, \ldots {X_n},} \right)$, ${Y} = \left( {{Y_1},{Y_2}, \ldots {Y_n},} \right)$, and ${Z} = \left( {{Z_1},{Z_2}, \ldots {Z_n},} \right)$
 of a shared random resource by Alice, Bob and Eve, respectively. We assume a scheme with unlimited message exchanges between Alice and Bob (available to Eve) whereby for some sufficiently large $n$, keys computed by Alice and Bob ($K_A$ and $K_B$, respectively) are made to satisfy the following requirements for some $\epsilon>0$,
 $\rm {(i)\ } \Pr \left( {{K_A} \ne {K_B}} \right) \le \varepsilon$,
 $\rm {(ii)\ } {n^{ - 1}}I\left( {{K_A};Z} \right) \le \varepsilon$,
 $\rm {(iii)\ } {n^{ - 1}}I\left( {{K_A}} \right) \ge {r_K} - \varepsilon$, and
 $\rm {(iv)\ } {n^{ - 1}}\log \left| C \right| \le {n^{ - 1}}H\left( {{K_A}} \right) + \varepsilon$,
where $H\left(  \cdot  \right)$
is the entropy, $I\left( { \cdot {\rm{ ;}} \cdot } \right)$
 is the mutual information between two random variables,   $I\left( { \cdot {\rm{ ;}}\left.  \cdot  \right| \cdot } \right)$
 is the mutual information between two random variables  conditioned on another random variable, $\left| C \right|$ is the cardinality of the key's alphabet ($C$), and ${r_K}$ is an achievable secret key rate of the scheme.
 In general, the secret key rate is not known, but an upper limit can be given by \cite{mau,ah}
${r_K} \le \min \left( {I\left( {X;Y} \right),I\left( {X;\left. Y \right|Z} \right)} \right)$,
where the mutual information between $X$ and $Y$ conditioned on $Z$ can be written
$I\left( {X;\left. Y \right|Z} \right) = H(X,Z) + H(Y,Z) - H(Z) - H(X,Y,Z)$, where $H\left( { \cdot , \cdot , \ldots } \right)$ is the joint entropy of a sequence of random variables.\footnote{Here,  $ I{(X;\left. Y \right|Z}) \le I({X;Y})$ for all our calculations.}
If we introduce the Kullback-Leibler information divergence between two probability mass functions $p(w)$ and $q(w)$ with sample space $W$, viz.,
$D\left( {\left. p \right\|q} \right) = \sum\limits_{w \in W} {p(w)\log \frac{{p(w)}}{{q(w)}}}  = {\rm E}\left[ { - \log \frac{{q(w)}}{{p(w)}}} \right]$ then the conditional mutual information can be estimated via a discrete formulation, viz.,
$
I\left( {X;\left. Y \right|Z} \right) = D\left( {\left. {p(x,y,z)} \right\|\frac{{p(x,z)p(y,z)}}{{p(z)}}} \right)$,
$ p(x)$ being the probability $X=x$, and $p\left( { \cdot , \cdot }, \ldots \right)$
   the corresponding joint probability.

\begin{figure}[!t]
	\begin{center}
		 {\includegraphics[width=3.4in,height=3.4in,clip,keepaspectratio]{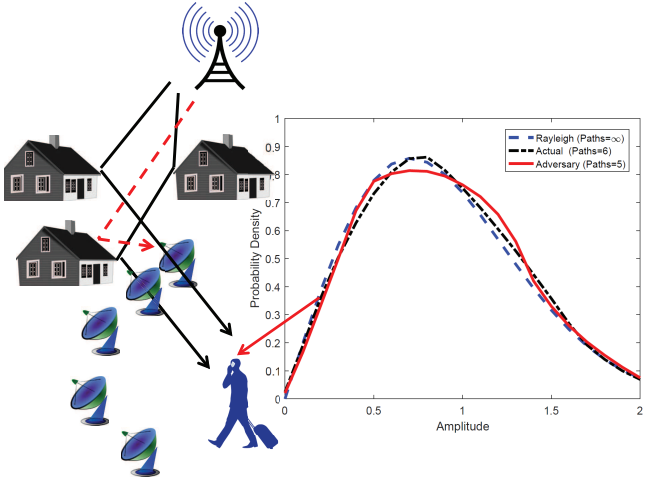}}
		\caption{Probability density functions (pdf) for different path settings. The inset graph shows the pdf at Bob for an infinite number of paths (dashed), 6 paths (dot-dashed), and a colluding Eve who has perfectly intercepted (see text) 5 paths (solid). The sketch on left illustrates the nature of the colluding attack where the solid (black) lines indicate some of the potentially many rays towards Bob that Eve intercepts, and the (red) dashed line indicates one of the potentially many `interference' paths to a directional antenna held by Eve.}\label{fig1}
	\end{center}
\end{figure}

We adopt the narrow-band flat-fading channel, and take the far-field approximation with wave propagation confined to a plane geometry. We assume the electric field vector is orthogonal to the plane and that isotropic gain antennas are held by Alice and Bob. If we consider, at the carrier frequency $f_c$ (wavelength $\lambda_c$), a bandpass transmitted signal $s(t) = {\mathop{\rm Re}\nolimits}\left\{ {\tilde s(t){e^{j2\pi {f_c}t}}} \right\}$, where $\tilde s(t)$ is the complex envelope,  the received bandpass signal can then be written (e.g. \cite{stub}), $r(t) = {\mathop{\rm Re}\nolimits} \left\{ {\sum\limits_{n = 1}^N {{C_n}{e^{j2\pi \left( {\left[ {{f_c} + f_n^D} \right]\left[ {t - {\tau _n}} \right]} \right)}}\tilde s(t - {\tau _n})} } \right\}$. Here $N$ is the number of propagation paths reaching the receiver, $C_n$ and $\tau_n$ are the amplitude and time delay, respectively,  and 
 $f_n^D$ is the Doppler frequency ($n$ indicates the $n$th path). This latter quantity can be expressed as $f_n^D
 =(v/\lambda_c)\cos {\alpha _n}$, $v$ being the velocity of the receiver and $\alpha_n$ being the angle of arrival (AoA) of the $n$th path at the receiver, relative to the velocity vector. Similar to the transmit signal, a complex envelope for the received signal can be written, $\tilde r(t) = \sum\limits_{n = 1}^N {{C_n}{e^{ - j{\varphi _n}(t)}}\tilde s(t - {\tau _n})}$, where ${\varphi _n}(t) = 
2\pi \left( {\left[ {{f_c} + f_n^D} \right]{\tau _n} - f_n^Dt} \right)$. Therefore, we have $r(t) = {\mathop{\rm Re}\nolimits} \left\{ {\tilde r(t){e^{j2\pi {f_c}t}}} \right\}$.
In the case  of a transmitted single tone this
can be written as $r(t) = {r_I}(t)\cos 2\pi {f_c}t - {r_Q}(t)\sin 2\pi {f_c}t$,
where
${r_I}(t) = \sum\limits_{n = 1}^N {{C_n}(t)\cos } {\varphi_n}(t)$  and ${r_Q}(t) = \sum\limits_{n = 1}^N {{C_n}(t)\sin } { \varphi_n}(t)$.    In the Rayleigh channel these quadratures  are independent Gaussian processes. Writing $\left| A \right| = \sqrt {{r_I}{{(t)}^2} + {r_Q}{{(t)}^2}}$ and $\vartheta  = \arctan \left( {{r_Q}(t)/{r_I}(t)} \right)$ we then have
$r(t) = \left| A \right|\cos \left( {2\pi {f_c}t + \vartheta } \right)$, where $\left| A \right|$ is Rayleigh distributed and $\vartheta$ is uniformly distributed.  In such a channel,  $\left| A \right|$ and/or $\vartheta$ can be used for secret key construction.

Ultimately the secret key is dependent on movement in the scattering environment, and it this movement that sets the channel coherence time (and therefore the key rate). The movement within the scattering environment ultimately manifests itself at the receiver through variation in received amplitudes and delays.
However, to enable clarity of exposition, we will make some simplifications to our scattering model - noting that the attack we describe can in fact be applied to any scattering environment scenario. The simplifications are that
we assume equal amplitudes for all received signals, and adopt random uniform distributions for all AoA and all phases as measured by the receiver. This is in fact 
the celebrated  2D isotropic scattering model of Clarke \cite{clark}. 
 Moving to the baseband signal henceforth for convenience, we note in Clarke's model the  signal of a transmitted symbol can be written as
$g(t) = \frac{1}{{\sqrt N }}\sum\limits_{n = 1}^N {\exp \left( {j\left( {{w_d}t\cos {\alpha _n} + \phi_n} \right)} \right)} ,
$
where  ${w_d}$ is  now the maximum Doppler frequency in rad/s, and $\phi_n$ is the  phase of  each path. Assuming both $\alpha_n$ and $\phi_n$ are independent and uniformly distributed in $\left[ { - \pi ,\pi } \right)$, then in the limit of large $N$  the amplitude of the signal $g(t)$ is distributed as a Rayleigh distribution.

Of particular interest to us here will be the statistics of $g(t)$ at low values of $N$, since in such circumstances the potential for an adversary to intercept all  signals is  larger. The higher order statistics  of the distribution within Clarke's model at finite $N$ have been explored in \cite{sinoid}, showing that  the following autocorrelation functions are in place; ${\Upsilon _{gg}}\left( \tau  \right) = {J_0}\left( {{\omega _d}\tau } \right)$, and ${\Upsilon _{{{\left| g \right|}^2}{{\left| g \right|}^2}}}\left( \tau  \right) = 1 + J_0^2\left( {{\omega _d}\tau } \right) - \frac{{J_0^2\left( {{\omega _d}\tau } \right)}}{N}$, where $J_0\left(  \cdot  \right)$ is the zero-order Bessel function of the first kind. For large $N$ these functions approach those of the Rayleigh distribution. Importantly,  the ${\Upsilon _{{{\left| g \right|}^2}{{\left| g \right|}^2}}}$ function is well approximated by an exponentiated sinc function at values of $N\ge 6$, meaning that (as per the usual assumption for any fading generated key), Eve must be several wavelengths away from Bob for the secret key rate to be non-zero.\footnote{A relaxation of this requirement may be obtained in specific correlated channel scenarios applicable to a distance of order 10 wavelengths ($\sim$ meters at GHz frequencies) away from the receiver\cite{He}. The attack  we describe here is unrelated to the special case of correlated channels. It is a general attack. 
 Eve's receivers can in principal be positioned anywhere (e.g. kms away from the intended receiver) yet still mount a successful attack.}

Many refinements on Clarke's model exist with perhaps the most widely used being that of \cite{sinoid}  
in which the main differentiator is a constraint placed on
the AoA, viz.,   
${\alpha _n} = \frac{{2\pi n + {\theta _n}}}{N}$, where $\theta_n$ is independently (relative to $\phi_n$) and uniformly distributed in $\left[ { - \pi ,\pi } \right)$. This latter simulator is wide sense stationary, is more efficient, and has improved second-order statistics.
In consideration of \textit{all} statistical measures, it is noted  that for this refined model any differences between  $N\gtrsim 8$  and the $N=\infty$ model (pure Rayleigh distribution) are largely inconsequential \cite{sinoid}.\footnote{In the calculations to follow, we find that the  key rates computed are, in effect, independent of whether this refined model  or Clarke's original model  is the adopted Rayleigh simulator.}

 In real-world channels, therefore, we have to be aware that even in cases where the channel appears to be consistent with a Rayleigh channel, the number of propagation paths 
 contributing  to the received signal can be relatively small. This can be seen more clearly from Fig.~(\ref{fig1}) where the probability density functions formed from six and five propagation paths are shown in comparison to the infinite path limit. The five path model corresponds to a case where Eve is missing one of the propagation paths used to construct Bob's signal.
 For the cases shown the  Kullback-Leibler divergence between the Rayleigh distribution and the lower-path models is very small.

 Let us assume the communications obtained by Bob consist of the combined signals from $N$ last-scattering events. We are interested in determining the effect, on some secret key generation scheme, caused by Eve's interception of all (or some fraction of) the $N$ last-scattered paths received by Bob. We assume Eve has  $M_e>>N$ directional-antenna receivers, and has placed them at multiple locations with the aim of continuously intercepting all of the last-scattered  signals towards Bob with high probability.\footnote{Such a possibility can be  enhanced in some scenarios by additional actions on Eve's part. For example, a scenario in which Eve has conducted an \emph{a priori} ray-tracing measurement (or analysis) campaign between a given pair of  transmit and receive locations thereby obtaining probabilistic information on likely last scattering points (for that given pair of locations). Of course in the limit ${M_E} \to \infty $ her probability of intercepting all paths approaches one in any case.} We  assume that these locations are much greater than $\lambda_c$ from Bob.

 Beyond our assumption of 2D geometry, and that the amplitude of each last-scattered  ray entering any receiver is equal, we also assume that the number of paths reaching each of Eve's antennas is  equal to $N$.\footnote{As an aside, we find a doubling of this number of paths at each of Eve's detectors has negligible impact on the results. Also note, as the number of paths  reaching Eve approach infinity, the size of her aperture must be made to approach infinity for the attack to remain viable. Neither limits are ever in place of course.}  Extension of our analysis to cover these issues is cumbersome, but straightforward. To make our mathematical notation less unwieldy, we will artificially set $M_e=N$ in our equations, with the understanding that we are then simply ignoring all of Eve's devices which (at any given time) are not intercepting any scattered rays towards Bob. 
  For added focus, we will assume Eve uses circular apertures of diameter $d$ as  her directional receivers - the physics and  properties of which can be found elsewhere, e.g.  \cite{book2}. Eve configures her $n$th aperture at each location so as to  maximize signal gain for the signal directed by the last scatterer in the direction of Bob (i.e. the $n$th of $N$ rays reaching Bob is centered in the main lobe of Eve's $n$th aperture). In such circumstances the signal collected by Eve's $n$th receiver  can be approximated as,
\[g_c^n(t) = \frac{1}{{\sqrt N }}\left\{ \begin{array}{l}
 \exp \left( {j\left( {w_d^et + \phi _n^e} \right)} \right) +  \\
 \sum\limits_{k = 2}^N {\left[ \begin{array}{l}
 \exp \left( {j\left( {w_d^et\cos \alpha _k^e + \phi _k^e} \right)} \right) \times  \\
 \frac{{2\lambda_c {J_1}\left( {\frac{{\pi d}}{\lambda_c }\sin \left( {\beta _k^e} \right)} \right)}}{{\pi d\sin \left( {\beta _k^e} \right)}} \\
 \end{array} \right]}  \\
 \end{array} \right\}\]
 where the superscript $e$ applied to any previously used variable means it is  now applied to Eve (but same meaning),  where
  ${\beta _k^e}$  represents the angle between the $k$th propagation path (side lobe `interference' path) arriving at Eve's detector and  ray~$n$  (i.e. ${\beta _n^e}=0)$,  and where
  $J_1\left(  \cdot  \right)$ is the Bessel function of the first kind of order one. Note that the maximum Doppler shift ${w_d^e}$ on Eve's detector is included so as to cover  the general case.   However, for focus we will assume all of Eve's detectors are  stationary, and in the following always set ${w_d^e}=0$. To reduce the mathematical complexity further we have not included in our analysis an obliquity factor $\left( {1 + \cos \beta _k^e} \right)/2$, which makes  our calculations conservative  (i.e. results in higher key rates).

  Upon receipt of the signals $g_c^n(t)$ Eve will adjust the signals for the known distance offset between each detector and Bob, and the known motion of Bob. This entails a phase adjustment   at each detector which manifests itself  as an `adjusting' phase $\phi_a^n$.
   The combined adjusted signal obtained by Eve after such signal processing, can then be written as $g(t) = \sum\limits_{n = 1}^N {g_c^n} (t)\exp (j\phi _{_a}^n)$.

   Assuming Eve's different apertures intercept all paths that are received by Bob,  the above relations lead us to conclude that, in principle, by increasing her aperture size Eve can determine Bob's received signal to arbitrary accuracy. In practice this accuracy will be limited by any  receiver noise on Eve's antennas, and error due to imprecise location information on Bob. However, with regard to these accuracy limitations (which we include in our Monte Carlo simulations below), we note the following two points that  favor Eve.
 (i) Given her all-powerful status,  Eve can set her noise to be at the quantum-limit (quantum noise).
 (ii) Beyond any other means available to her, an unlimited  Eve can determine  the location of Bob at time $t$ to any required accuracy through signal acquisition. More specifically in regard to (ii),  the minimum position error via signal processing varies as $1/\sqrt {{M_e}}$ - a result that holds even if some of Eve's devices are affected by shadowing in which the path-loss exponents are unknown \cite{malaney2}.

To make further progress we must introduce an actual scheme for generating a secret key. Although there are many such schemes (e.g. \cite{surv1,surv2,rev2,poor})  we will adopt here a generic formulation that covers the conceptual framework of the widely used signal threshold schemes. The basic concept of such schemes is to quantize a received signal metric, say amplitude, into a 1 or 0 value. For some parameter $T>0$, and for some median value $m$ of the expected amplitude distribution, the decision value can then be set dependent on whether the  amplitude is below $m-T$ or above $m+T$.  Such schemes offer many pragmatic advantages and compensate to a large extent errors introduced through a lack of exact reciprocity between transceiver configurations. Assuming a given level of Gaussian noise at Bob and Eve's receivers, an appropriate  value of $T$ can be chosen. Further, so as to maximize the entropy of the final key, we introduce an  `entropy'   factor $s$.  For a given $T$ and probability density function $R'(r)$ for the received amplitude $r$,  the value of $s$  can be determined through $\int_0^{m-T} {R'(r)} dr = \int_{m +T+ s}^\infty  {R'(r)} dr$. Note, in general $R'$ is the distribution for the amplitudes in the presence of non-zero Gaussian receiver noise. When $r$ is measured by Alice and/or Bob to be  between the two `allowed' regions,  as defined by the integrals of this relation, it is agreed by both parties that the measurement be dropped.

 Clearly, in practice larger values of $T$ will minimize mismatches in the key at the cost of a reduced key generation rate. Ultimately, in any real scheme a period of reconciliation and privacy amplification will be pursued in order to obtain the final key. However, here we will simply investigate  the upper limit of the key rate through a numerical evaluation of the conditional mutual information as defined earlier. We assume Eve's strategy on detection is to decide on the binary number in the `disallowed' region by setting $s=T=0$. We also assume all issues on the decision strategy of the scheme and all communications between Alice and Bob (e.g. which measurements to drop) are available to Eve.

\begin{figure}[!t]
	\begin{center}
		{\includegraphics[width=3.7in,height=5.2 in,clip,keepaspectratio]{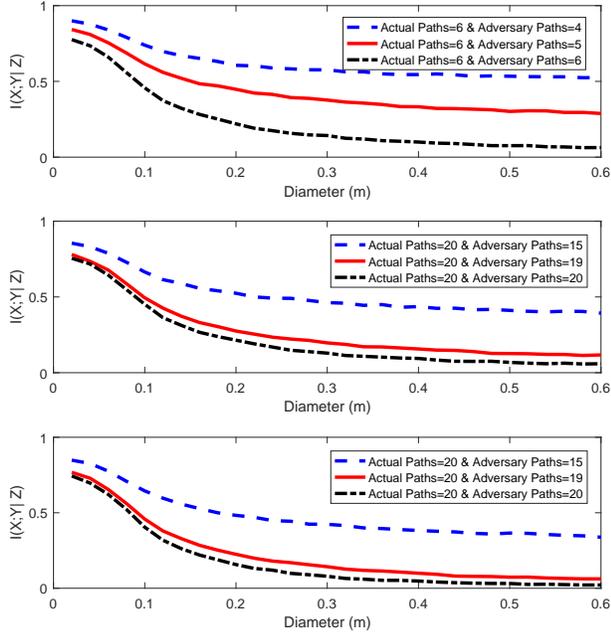}}
		\caption{Change in the conditional mutual information between Alice and Bob as function of the diameter of Eve's directional antenna (a circular aperture) for different path conditions. Six paths (top figure) and 20 paths (middle figure) are used to construct the approximate Rayleigh distribution. One calculation (bottom figure) on the 20 path scenario assumes zero receiver noise at Eve and zero location error on Bob. Results shown are for 1 million Monte Carlo runs.
}\label{fig2}
	\end{center}
\end{figure}

 Fig.~(\ref{fig2}) (top) displays a calculation of the conditional mutual information as a function of aperture diameter (all of Eve's circular apertures are assumed to be the same size) in which a receiver noise contribution (on all receivers) is set so that the signal-to-noise ratio (SNR) is equal to 17dB. The maximum Doppler shift of Bob is set to 10Hz, $\lambda_c$ is set to 0.1m, and a Gaussian error on the pointing of Eve's apertures (due to location error on Bob) is set to a standard deviation of 0.002 radians. The threshold is set at three times the receiver noise.
 We can see that if all signals are intercepted the key rate can be driven to almost zero over the range of aperture diameters probed.   For fewer signals intercepted we see that useful key rates are still possible, albeit at significantly diminished values relative to a no-attack scenario.   For comparison,  Fig.~(\ref{fig2}) (middle) displays similar calculations   but for 20 propagation paths forming the Rayleigh distribution, and  Fig.~(\ref{fig2}) (bottom) shows the same calculation when Eve's detectors are operating with zero receiver noise, and location errors on Bob are assumed to be zero.

 The specific key scheme discussed here is limited in scope relative to the large number of possible key generation schemes available. More sophisticated schemes, such as those based on multi-antenna transceiver configurations, the use of optimal coding techniques, and the use of channel state information, are possible.
 However, straightforward extensions of the attack described here would still apply to all of these more sophisticated schemes - only the quantitative details on how the key rate is diminished under the attack will be different.  
 
 Indeed, we note the attack described here can be generalized further so as to \emph{always} drive the secret key rate to zero, even if we relax the assumption that it is only the last-scattering rays that are intercepted. An all-powerful Eve, with ${M_E} \to \infty$, can intercept all propagation paths (of any energy) at all points in space, and in principal possess knowledge on all characteristics of all scatterers. With the unlimited computational resources afforded to her the classical Maxwell equations can then be solved exactly, thereby providing  information on any of Bob's received signals at an accuracy  limited only by quantum mechanical effects. Of course, such an  attack whilst theoretically possible, is not tenable. The calculations described here  can be considered a limited  form of such an attack, tenable
 in a real-world scattering environment.

 In conclusion, we have described a new attack on classical schemes  used to generate secret keys via the shared randomness inherent in wireless fading channels.
  Although the attack we have described will be difficult to implement in a manner that drives the secret key rate to zero, our work does illustrate  how  such a rate can at least be partially reduced.  As such, all schemes for secret key generation via the fading channel must invoke a new restriction - a limitation on the combined information received by a colluding Eve.



%

%
%

\end{document}